\documentclass[useAMS]{mn2e}

\usepackage{graphicx}

\title[IC~4663]{IC~4663: The first unambiguous [WN] Wolf-Rayet central star of a planetary nebula\thanks{Based on observations made with Gemini South under programme GS-2011A-Q-65 (PI: B. Miszalski).}}

\author[B. Miszalski et al.]{B. Miszalski,$^{1,2}$\thanks{E-mail: brent@saao.ac.za} P. A. Crowther,$^{3}$ O. De Marco,$^{4}$ J. K\"oppen,$^{5,6,7}$ A. F. J. Moffat,$^{8}$ \newauthor A. Acker$^{5}$ and T. C. Hillwig$^{9}$\\
$^{1}$South African Astronomical Observatory, PO Box 9, Observatory, 7935, South Africa\\
$^{2}$Southern African Large Telescope Foundation, PO Box 9, Observatory, 7935, South Africa\\
$^{3}$Department of Physics and Astronomy, Hounsfield Road, University of Sheffield, Sheffield S3 7RH, UK\\
$^{4}$Department of Physics and Astronomy, Macquarie University, Sydney, NSW 2109, Australia\\
$^{5}$Observatoire astronomique de Strasbourg, Universit\'e de Strasbourg, CNRS, UMR 7550, 11 rue de l'Universit\'e, F-67000 Strasbourg, France\\
$^{6}$International Space University, Parc d'Innovation, 1 Rue Jean-Dominique Cassini, F-67400, Illkirch-Graffenstaden, France\\
$^{7}$Institut f\"ur Theoretische Physik und Astrophysik, Universit\"at Kiel, D-24098, Kiel, Germany\\
$^{8}$D\'ept. de physique, Univ. de Montr\'eal C.P. 6128, Succ. Centre-Ville, Montr\'eal, QC H3C 3J7, and Centre de recherche\\ en astrophysique du Qu\'ebec, Canada\\
$^{9}$Department of Physics and Astronomy, Valparaiso University, Valparaiso, IN 46383, USA
}

\begin{document}

\date{Accepted Received ; in original form }

\pagerange{\pageref{firstpage}--\pageref{lastpage}} \pubyear{2012}

\maketitle

\label{firstpage}

\begin{abstract}
   We report on the serendipitous discovery of the first central star of a planetary nebula (PN) that mimics the helium- and nitrogen-rich WN sequence of massive Wolf-Rayet (WR) stars. The central star of IC~4663 (PN G346.2$-$08.2) is dominated by broad He~II and N~V emission lines which correspond to a [WN3] spectral type. Unlike previous [WN] candidates, the surrounding nebula is unambiguously a PN. At an assumed distance of 3.5 kpc, corresponding to a stellar luminosity of 4000 $L_\odot$, the $V=16.9$ mag central star remains 4--6 mag fainter than the average luminosity of massive WN3 stars even out to an improbable $d=8$ kpc. The nebula is typical of PNe with an elliptical morphology, a newly discovered Asymptotic Giant Branch (AGB) halo, a relatively low expansion velocity ($v_\mathrm{exp}=30$ km s$^{-1}$) and a highly ionised spectrum with an approximately Solar chemical abundance pattern. The [WN3] star is hot enough to show Ne~VII emission ($T_\ast=140\pm20$ kK) and exhibits a fast wind ($v_\infty$=1900 km s$^{-1}$), which at $d=3.5$ kpc would yield a clumped mass loss rate of $\dot{M}$ = 1.8 $\times 10^{-8} M_{\odot}$\,yr$^{-1}$ with a small stellar radius ($R_\ast=0.11$ $R_\odot$). Its atmosphere consists of helium (95\%), hydrogen ($<2$\%), nitrogen (0.8\%), neon (0.2\%) and oxygen (0.05\%) by mass. Such an unusual helium-dominated composition cannot be produced by any extant scenario used to explain the H-deficiency of post-AGB stars. The O(He) central stars share a similar composition and the discovery of IC~4663 provides the first evidence for a second He-rich/H-deficient post-AGB evolutionary sequence [WN]$\to$O(He). This suggests there is an alternative mechanism responsible for producing the majority of H-deficient post-AGB stars that may possibly be expanded to include other He-rich/H-deficient stars such as R Coronae Borealis stars and AM Canum Venaticorum stars. The origin of the unusual composition of [WN] and O(He) central stars remains unexplained.

\end{abstract}


   \section{Introduction}
   Planetary nebulae (PNe) are the result of low-intermediate mass stars ($\sim$1--8 $M_\odot$) that have experienced heavy mass loss during the asymptotic giant branch (AGB) phase. Their gaseous envelopes are ionised by hot pre-white dwarfs ($T_\mathrm{eff}$$\sim$30--140 kK) whose diverse spectroscopic appearance (M\'endez 1991) reflects different wind properties and atmospheric compositions. A subset of central stars have H-deficient atmospheres composed primarily of helium, carbon and oxygen with fast and dense winds that produce broad emission lines. Their spectroscopic appearance mimics the carbon-sequence of massive Wolf-Rayet (WR) stars and they are classified into either [WO] or [WC] types, the brackets distinguishing them apart from massive stars (van der Hucht 2001), depending on which emission lines are present and their relative intensities (Crowther, De Marco \& Barlow 1998; Acker \& Neiner 2003). Despite recent surveys that have dramatically increased the number of central stars that belong to the [WC] sequence (Acker \& Neiner 2003; G\'orny et al. 2004, 2009, 2010; DePew et al. 2011), their origin and evolution remain poorly understood.

   Werner \& Herwig (2006) reviewed possible scenarios that may explain the H-deficiency of [WC] central stars\footnote{Massive WC stars are extremely H-deficient, but for a different reason. A strong wind has peeled off the outer H-containing layers exposing in turn the C- and O-rich products of triple-$\alpha$ He-burning.} and their progeny the PG1159 stars (Wesemael, Green \& Liebert 1985). After the central star has left the AGB phase a late thermal pulse (LTP) or very-late thermal pulse (VLTP) would reignite helium-shell burning (e.g. Bl\"ocker 2001; Herwig 2001). The star is sent back to the AGB phase, i.e. it becomes ``born-again'' (Sch\"onberner 1979), and any residual hydrogen is either burned or diluted below the observable detection limit. The AGB final thermal pulse (AFTP) variant may also be invoked to explain some instances where a small amount of observable hydrogen remains. At this point the star enters the [WC]-sequence with an H-deficient atmosphere mostly made of helium ($\sim$30--50\%), carbon ($\sim$30--60\%) and oxygen ($\sim$2--20\%).
   
   A post-AGB H-deficient evolutionary sequence [WCL]$\to$[WCE]$\to$PG1159 has been proposed in which late-type [WC] central stars evolve into early-type [WC] and [WO] central stars, followed by PG1159 stars which appear preferentially in the oldest PNe (Napiwotzki \& Sch\"onberner 1995) and in isolation once the surrounding PNe have dissipated into the interstellar medium.\footnote{This differs from massive WC stars which appear to have their subtype determined by their initial metallicity, although some evolution from late to early may also occur (Smith \& Maeder 1991).} Unless there is a residual amount of hydrogen the end product will be a non-DA white dwarf. This evolutionary sequence is supported by the strong similarities in derived atmospheric compositions (Werner \& Herwig 2006) and some transition objects (e.g. Bond 2008). A key question is whether the H-deficiency of all members of the sequence can be explained by the LTP, VLTP and AFTP scenarios. G\'orny \& Tylenda (2000) found the evolutionary status of nebulae surrounding H-deficient central stars to be consistent with H-rich central stars, rather than following the paths predicted by LTP or VLTP scenarios. It is therefore debatable whether the few documented LTP and VLTP events (Sch\"onberner 2008) mandate the same origin for the larger population of H-deficient central stars. Indeed, there may even be an alternative common-envelope (De Marco \& Soker 2002; De Marco 2008; Hajduk, Zijlstra \& Gesicki 2010) or nova-like explanation for such objects (Wesson, Liu \& Barlow 2003; Wesson et al. 2008; Lau, De Marco \& Liu 2011).
   
   Even more difficult to explain are the helium-rich H-deficient stars which do not fit the carbon-rich post-AGB H-deficient sequence. These include four O(He) stars (Rauch et al. 1998), two of which are central stars of PNe (Rauch et al. 1994, 1996), R Coronae Borealis (RCB) stars (Clayton 1996), extreme helium B stars and helium-rich subdwarf O stars (see e.g. Werner \& Herwig 2006). With typical helium mass fractions $\ga$90\% and trace amounts of other elements the chemical compositions of these stars have never been predicted by the born-again scenarios, i.e. they simply should not exist! On the whole these stars are scarce compared to the carbon-rich sequence and probably form a second, parallel post-AGB H-deficient sequence. Evidence for such a sequence is tentative however, based primarily on their common composition rather than other evolutionary ties. It has been suggested that RCB stars are the progenitors of O(He) stars, although the circumstances of this possible connection remain unclear. After the discovery of a low $^{16}$O/$^{18}$O ratio by Clayton et al. (2007) strongly pointing to a merger origin for RCB stars, an end to the protracted debate on the formation of RCB stars seemed to be in sight. However, not only is the RCB evolutionary path complicated by a connection with [WC] central stars (Clayton \& De Marco 1997; Clayton et al. 2006; Clayton et al. 2011a), but there are indications of a mixed origin within the RCB class itself (Clayton et al. 2011b).

   It is imperative that more members of the He-rich/H-deficient class are found and studied to establish a second evolutionary sequence. If this can be achieved, then there should exist another mechanism responsible for H-deficiency, one which may also help explain the origin of the carbon-rich sequence. A long sought after solution may be [WN] type WR central stars that mimic the helium- and nitrogen-rich WN sequence in massive stars (Smith, Shara \& Moffat 1996). Although some candidates have been proposed, none have turned out to be satisfactory and the existence of the class remains unproven.  
   
   Here we report on the serendipitous discovery of a [WN] central star in the PN IC~4663 (PN G346.2$-$08.2, Fleming \& Pickering 1910). Observations of IC~4663 were made during the Gemini South programme GS-2011A-Q-65 (PI: B. Miszalski) which aimed to discover new close binary central stars of PNe via radial velocity variability (e.g. Miszalski et al. 2011a). Program targets were selected based on the nebula trends identified by Miszalski et al. (2009b) to be provisionally associated with binarity that includes bipolar nebulae, nebulae with collimated outflows (jets) and low-ionisation structures (e.g. Gon{\c c}alves et al. 2001). 
   As noted by Miszalski et al. (2009b), a large number of PNe with WR central stars overlap with this morphological prescription and so it was not unexpected that new examples would be found during the programme.

This paper is structured as follows. Section \ref{sec:prev} reviews previous [WN] candidates. Section \ref{sec:cspn} describes Gemini South spectroscopy and atmospheric modelling of the previously unobserved central star of IC~4663. The basic nebula properties and chemical abundances are presented in Section \ref{sec:nebula}. Section \ref{sec:discussion} establishes the [WN] nature of the central star by proving the bona-fide PN nature of IC~4663 and introduces the second post-AGB H-deficient evolutionary sequence. We conclude in Section \ref{sec:conclusion}.

   \section{Previous [WN] candidates}
   \label{sec:prev}
\begin{figure}
   %Throughput of STIS clear filter can be found in Fig 5.1. of STIS manual.
      \begin{center}
         \includegraphics[scale=0.3]{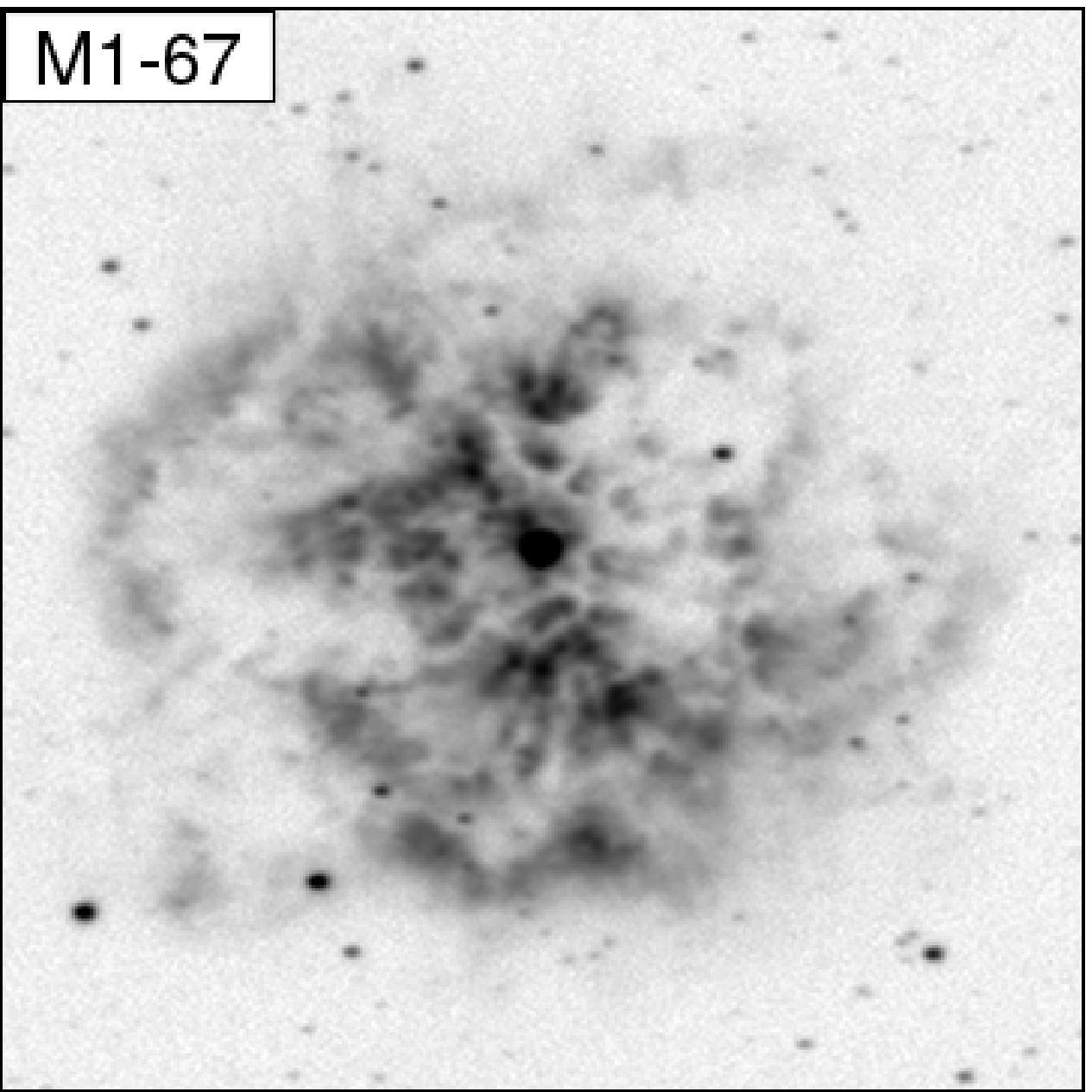}
         \includegraphics[scale=0.3]{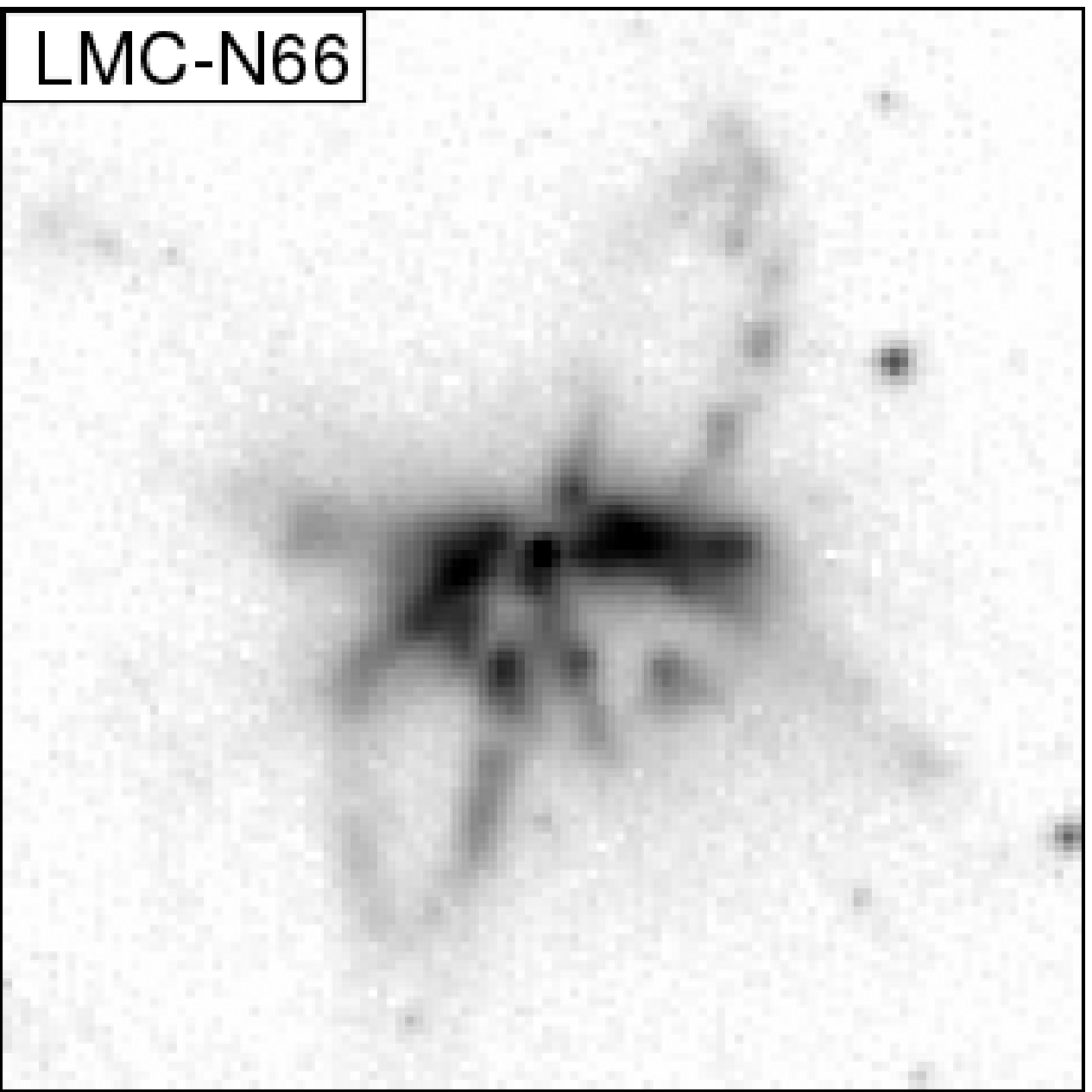}\\
         \includegraphics[scale=0.3]{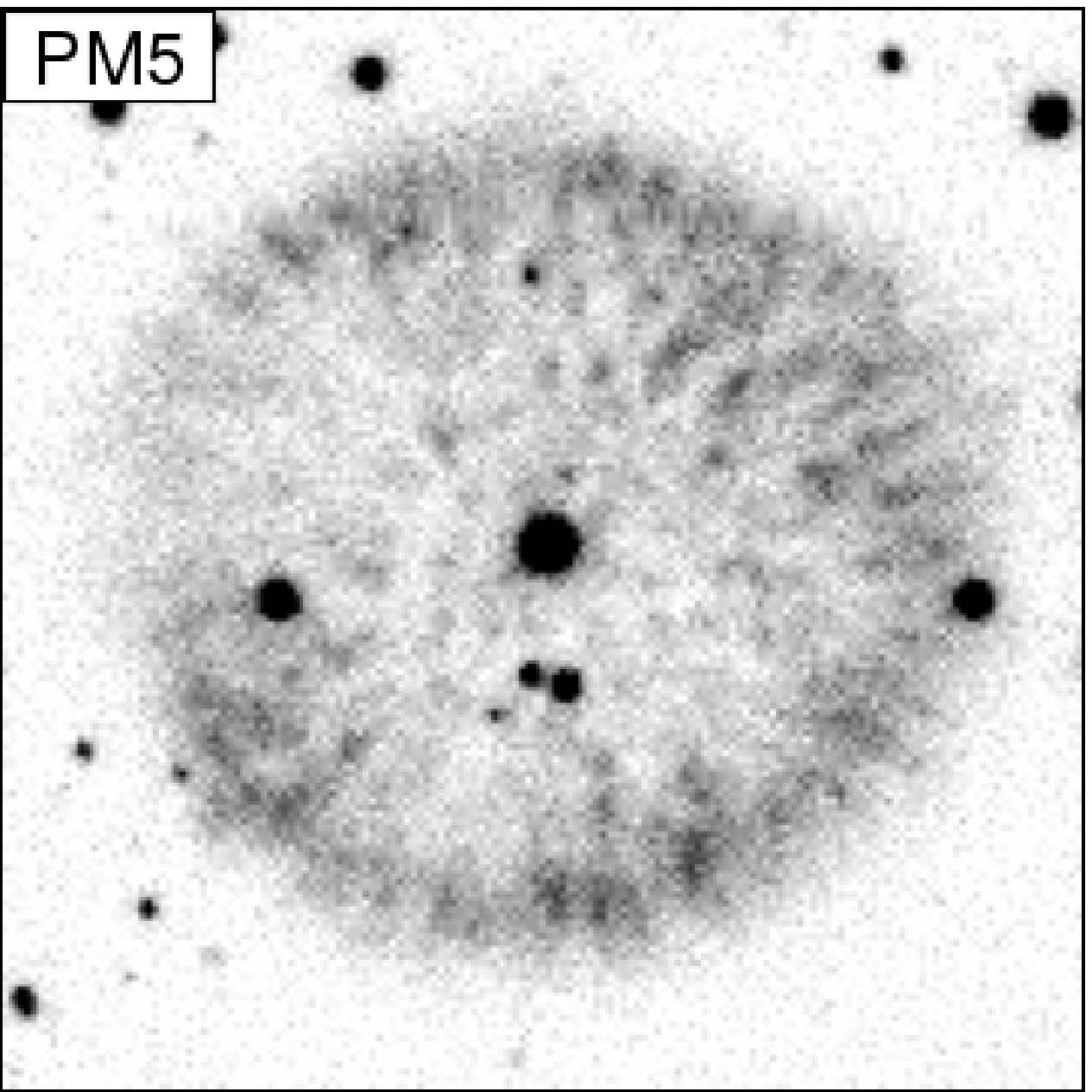}
         \includegraphics[scale=0.3]{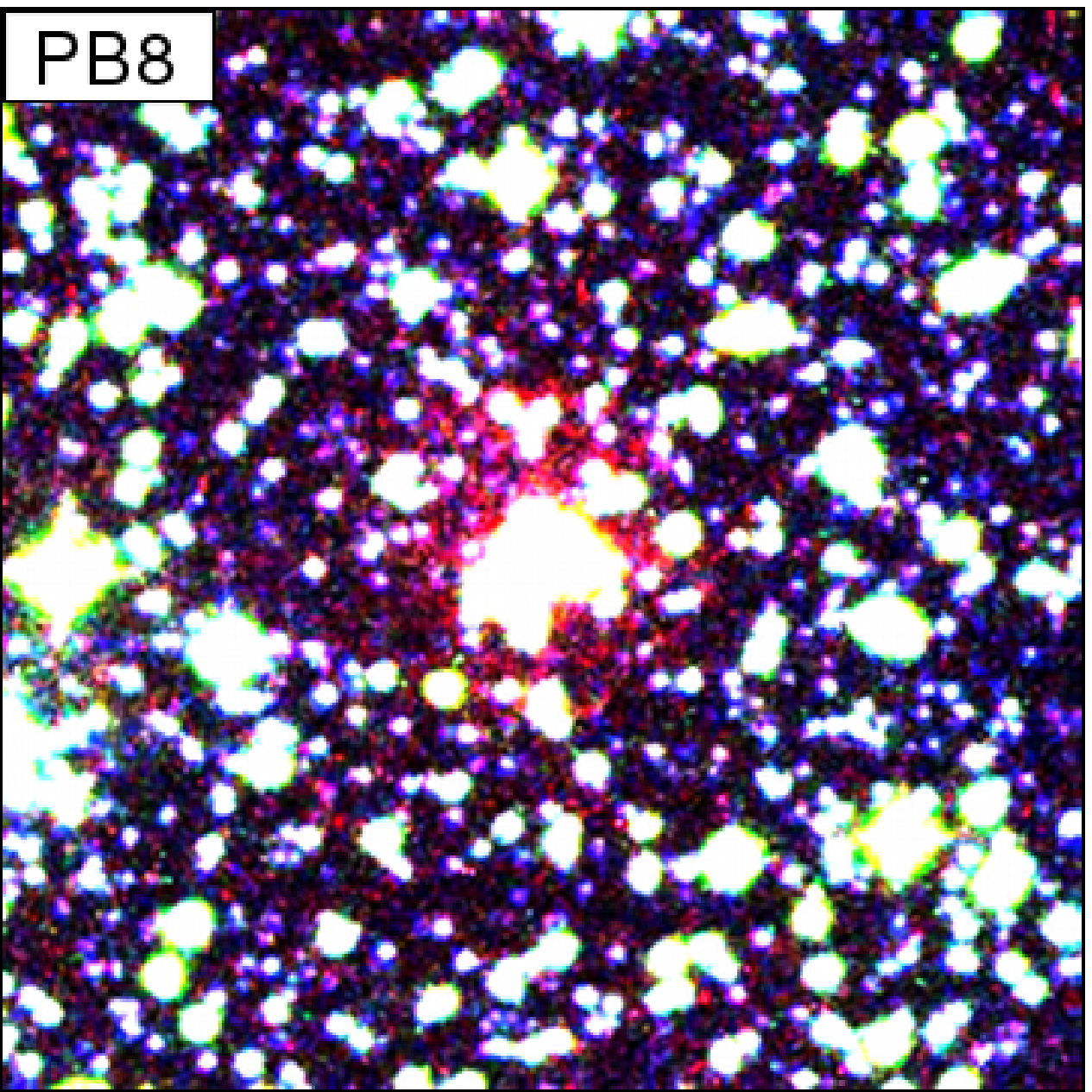}
      \end{center}
      \caption{A selection of previous [WN] candidates. The M~1-67 (Drew et al. 2005; Gonz\'alez-Solares et al. 2008) and PM~5 (see text) images were both taken with an H$\alpha$+[N~II] filter, while LMC-N66 was observed with the clear filter of \emph{HST} STIS.
The colour-composite image of PB~8 is made from SHS H$\alpha$+[N~II] (red), SHS Short-Red (green) and SuperCOSMOS Sky Survey $B_J$ (blue, Hambly et al. 2001). It reveals for the first time a faint AGB halo with a diameter up to $\sim$80\arcsec\ surrounding PB~8 (the bright central source). Image dimensions are $120\times120$ arcsec$^2$ (M~1-67), $6\times6$ arcsec$^2$ (LMC-N66), $40\times40$ arcsec$^2$ (PM~5) and $4\times4$ arcmin$^2$ (PB~8). In all cases North is up and East to left. }
      \label{fig:imgs}
   \end{figure}

   Around 8\% of massive Galactic WN stars are known to have surrounding nebulae (Stock \& Barlow 2010). Until improved observations could reliably determine the luminosity of their central stars, some of these nebulae were misclassified as PNe. As the distances to PNe are extremely difficult to estimate this has not helped clarify the status of many candidates. 
   Figure \ref{fig:imgs} shows M~1-67 which was included in the PN catalogue of Perek \& Kohoutek (1967), however its WN8h star WR124 was only later proven to be a massive star by Crawford \& Barlow (1991a) based on the kinematic distance measured from the interstellar Na I D$_2$ absorption line.\footnote{Marchenko, Moffat \& Crowther (2010) recently confirmed this result by analysing the nebula expansion parallax with 2-epoch \emph{HST} observations.} The flocculent appearance of the nebula, evident in Fig. \ref{fig:imgs}, but best seen in \emph{HST} observations (Grosdidier et al. 2001a), also bears more resemblance to nebulae around massive WN stars including RCW~58 around WR40 (Chu 1982) and NGC6888 around WR136 (Wendker et al. 1975). Another similar case to M~1-67 is We~21 (Duerbeck \& Reipurth 1990) which was proven to host a massive star by Crawford \& Barlow (1991b).

   Other [WN] candidates have proven very difficult to establish as promising candidates for clear-cut low-mass analogues of the massive WN sequence. The most extensively studied is LMC-N66 in the Large Magellanic Cloud (see e.g. Pe\~na et al. 1994, 1997, 2004, 2008; Hamann et al. 2003). Multiple outbursts have been recorded in the central star which shows a mid-WN spectral type during outburst. Pe\~na et al. (2004) studied the surrounding nebula (Fig. \ref{fig:imgs}) which resembles point-symmetric PNe in the Milky Way (e.g. Fleming 1, L\'opez et al. 1993) and shows multiple emission components that span 130--160 km s$^{-1}$. The luminosity spans log $L/L_\odot=$4.6--5.4 (Hamann et al. 2003), too high to be consistent with a typical PN central star. Hamann et al. (2003) presented a number of possible explanatory scenarios, none of which are without difficulty, although a massive star in a common-envelope phase with a low mass companion or a massive white dwarf (WD) accreting matter in a close binary seem to be preferred. Binary stellar evolution is therefore likely responsible for the WN features of LMC-N66, as is similarly seen in symbiotic novae such as RR Tel and PU Vul (Thackeray \& Webster 1974; Nussbaumer 1996). 
   
   Morgan, Parker \& Cohen (2003) discovered a WN6 star in the nebula PM5 and argued for a PN status based on the nebula density, size, ionised mass, morphology, near-infrared and mid-infrared properties mostly being similar to PNe. A non-PN status is supported by the anomalously high expansion velocity (165 km s$^{-1}$), low Galactic latitude ($b=+0.69$ deg), high reddening ($E(B-V)$\ =\ 3.0--3.5 mag) and the analysis by Todt et al. (2010b). Furthermore, Fig. \ref{fig:imgs} shows a previously unpublished VLT FORS2 (Appenzeller et al. 1998) image observed under ESO programme ID 077.B-0430(B) (PI: M. Pe\~na). The 20 sec image taken through the H\_alpha$+$83 filter (central wavelength ($\lambda_\mathrm{cen}$) and full-width at half-maximum (FWHM) of 656.3 nm and 6.1 nm, respectively) reveals a flocculent appearance which was not visible in the lower resolution SuperCOSMOS H$\alpha$ Survey discovery images (SHS, Parker et al. 2005). 
   Similarly, the WN6 central star of A~48 (Abell 1966; Wachter et al. 2010; DePew et al. 2011), is also likely to be a massive star, however this object has not yet been studied in detail. It is also located at a low Galactic latitude ($b=+0.4$ deg), has a double-ring morphology (Stock \& Barlow 2010) and is also quite reddened ($E(B-V)=2.0$ mag, DePew et al. 2011). 
   
   Several other candidates have also been identified based on the presence of weak nitrogen emission features in addition to their [WO/WC] spectra. The [WO4/WC4] central stars of NGC~6751 (Aller 1968; Koesterke \& Hamann 1997), PMR1 and PM3 (Morgan, Parker \& Russeil 2001; Parker \& Morgan 2003) exhibit weak N~IV and N~V emission features. Further examples are evident amongst the [WO4pec] and [WO] central stars studied by Acker \& Neiner (2003). In the case of NGC~6751 nitrogen accounts for no more than 1\% by mass in an otherwise typical [WO/WC] atmospheric composition of 55\% helium, 30\% carbon and 15\% oxygen (Koesterke \& Hamann 1997). Others were analysed by Todt et al. (2008) who found nitrogen contributed no more than 1--2\% by mass.
   
   Todt et al. (2010a, 2010b) found the central star of PB~8 to show clear nitrogen emission features and determined a nitrogen mass fraction of 2\%. Its overall atmospheric composition was found to be made of oxygen and carbon each 1.3\% by mass, helium 55\% and hydrogen an unusually high 40\%. This is most unlike [WO/WC] central stars which have significantly higher fractions of carbon and oxygen (Crowther 2008). The PN nature of PB~8 is demonstrated by its relatively low expansion velocity ($19\pm6$ km s$^{-1}$) and especially our discovery in Fig. \ref{fig:imgs} of a faint AGB halo (Corradi et al. 2003). Todt et al. (2010a) proposed that PB~8 belongs to a [WN/WC] classification, an analogue of the massive WN/WC stars (e.g. Massey \& Grove 1989; Crowther et al. 1995a) which account for $<$5\% of massive WR stars (van der Hucht 2001). Therefore, strictly speaking, PB~8 is not a true [WN] nucleus due to the additional [WC] features. 

   To summarise, no [WN] candidate has been unambiguously proven to be surrounded by a PN. While PB~8 is certainly a PN, it appears to be a [WN/WC] hybrid rather than a strict [WN].
   
   \section{The central star}
   \label{sec:cspn}
   \subsection{Observations}
   We used the Gemini Multi-Object Spectrograph (GMOS, Hook et al. 2004) on Gemini South to obtain three spectra of the central star of IC~4663 on 27 June, 2 July and 3 July 2011 as part of the GS-2011A-Q-65 programme (PI: B. Miszalski). The B1200 grating was used with a 0.75\arcsec\ slit oriented at a position angle of 90$^\circ$ to provide wavelength coverage from 3767--5228 \AA\ (27 June and 3 July) and 4507--5977 \AA\ (2 July) at 1.6 \AA\ resolution (FWHM) with a reciprocal dispersion of 0.23 \AA\ pixel$^{-1}$ ($1\times2$ binning in dispersion and spatial axes, respectively). The science exposure times were 1800 s (27 June) and 2400 s (2 and 3 July) and were accompanied by contemporaneous arc lamp exposures. The data were reduced with the Gemini \textsc{iraf} package after which nebula-subtracted one-dimensional spectra were extracted using the \textsc{apall} \textsc{iraf} task. A non-uniform spatial profile in the nebula emission lines produced a small amount of over-subtraction at the Balmer lines, He~II $\lambda\lambda$4686, 5412, and He~I $\lambda$5876. The one-dimensional spectra were flux calibrated with a spectrum of the spectrophotometric standard star EG274 in the usual fashion.

   The $V$ magnitude of the central star is not well determined in the literature with discordant estimates of $V=15.16^{+0.29}_{-0.23}$ mag (Shaw \& Kaler 1989) and $B=18.0\pm$(0.5--1.0) mag (Tylenda et al. 1991). We retrieved the processed and combined \emph{HST} $F555W$ image (approx. Johnson $V$) of total duration 1020 s from the Hubble Legacy Archive\footnote{http://hla.stsci.edu} (HLA) and performed aperture photometry on the central star following the WFPC2 Photometry Cookbook.\footnote{http://www.stsci.edu/hst/wfpc2/analysis/wfpc2\_cookbook.html} Assuming $V-F555W=0.05$ mag, suitable for an O5V star, we found $V=16.90\pm0.05$ mag and fixed the absolute flux scale of our GMOS spectrum to match the corresponding Vega flux at $\lambda$5480 \AA. 

   \subsection{Spectral features and [WN3] classification}
\begin{figure*}
      \begin{center}
         \includegraphics[scale=0.75,angle=270]{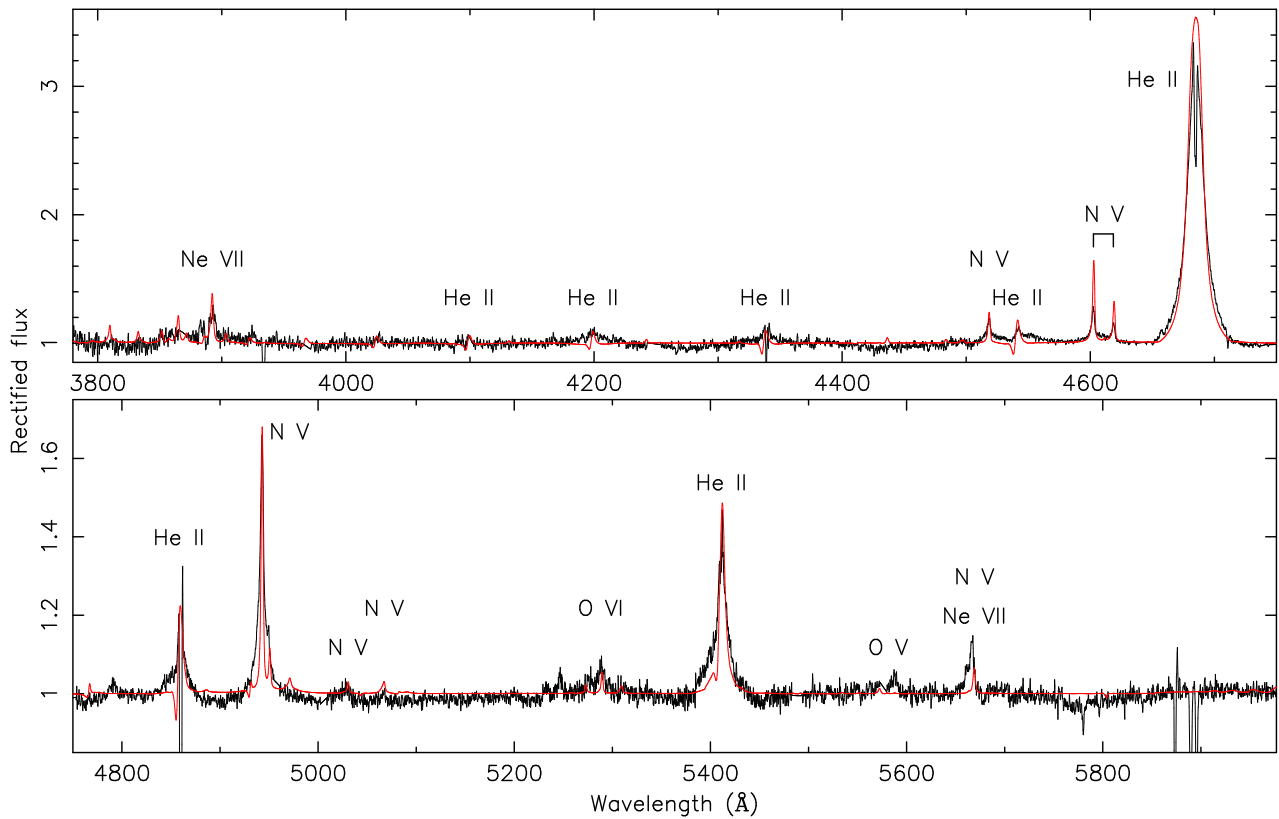}
      \end{center}
      \caption{Rectified Gemini South GMOS spectrum of IC~4663 (black) with the \textsc{cmfgen} NLTE model (red, see Sect. \ref{sec:nlte}). The main emission lines are identified, most belonging to He~II and N~V, which together with the absence of C~IV and He~I emission, indicates a [WN3] spectral type. The apparent double-peaked profile of He~II $\lambda$4686 is an artefact of nebula subtraction and does not influence the NLTE analysis.}
      \label{fig:spec}
   \end{figure*}

   Figure \ref{fig:spec} shows the two 2400 s exposures averaged to achieve a signal-to-noise ratio (S/N) of at least $\sim$60 in the continuum (except for $\lambda\le4500$ \AA\ where S/N $\ga30$) and rectified for comparison with the NLTE model described in Sect. \ref{sec:nlte}. The spectrum is dominated by He~II and N~V emission lines typical of early-type WN stars (Crowther et al. 1995b) and shares the triangular line profiles of Galactic early-type WN stars (Smith et al. 1996). The Ne~VII emission lines (Werner, Rauch \& Kruk 2007) were never before seen in massive WN stars, whereas the weak O~V and O~VI emission are uncommon (e.g. the WN3pec star WR46 shows relatively strong O~VI $\lambda$3811-34, Crowther et al. 1995b). The high S/N spectrum rules out the presence of He~I and C~IV emission lines which, combined with the already mentioned features, corresponds to a clear WN3 subtype (Smith et al. 1996). The WN3 subtype is not common amongst massive Galactic WN stars where only a few examples are known (e.g. Crowther et al. 1995b; Hamann et al. 1995; Smith et al. 1996). The equivalent width of He~II $\lambda$4686 ($\sim$40 \AA) is comparable to massive WN3 stars, whereas its FWHM is slightly lower than WN3 stars ($\sim$18.5 \AA). This behaviour is also seen in C~III $\lambda$5696 in the comparison between the [WC9] central star of BD+30~3639 and massive WC9 stars (Crowther et al. 2006). In the rest of this paper we denote the classification as [WN3] to indicate it is a bona-fide central star of a PN (later proven in Sect. \ref{sec:bona}).\footnote{Despite the good match with WN3 stars, we caution that the Smith et al. (1996) scheme was solely devised for massive WN stars. If a sufficient sample of [WN] stars were to be found, then a unified classification scheme may be required (e.g. Crowther et al. 1998) which may not necessarily match the existing one.} 

\subsection{NLTE modelling}
\label{sec:nlte}

We have analysed the GMOS spectrum of IC 4663 using \textsc{cmfgen} (Hillier \& Miller 1998) which solves the radiative transfer equation in the comoving frame subject to statistical and radiative equilibrium, assuming an expanding, spherically symmetric, homogeneous atmosphere, allowing for metal line blanketing and wind clumping.

Our adopted model atom includes H\,{\sc i}, He\,{\sc i-ii}, C\,{\sc iv}, N\,{\sc iv-v}, O\,{\sc iv-vi}, Ne\,{\sc iii-viii}, Si\,{\sc iv} and Fe\,{\sc v-ix}. In total 512 superlevels, 1907 levels, and 28397 non-local thermodynamic equilibrium (NLTE) transitions are simultaneously considered, with H/He, C/He, N/He, O/He variable, and Si and Fe mass fractions fixed at Solar values (Asplund et al. 2009). We adopt a Ne mass fraction of 0.2\%, based on a number ratio of Ne/O = 0.25 from Wang \& Liu (2008) and the Solar O mass fraction of 0.55\% (Asplund et al. 2009).

In view of the relatively weak emission line spectrum of IC 4663, the supersonic velocity law, $v_{\infty}(1-R/r)^{\beta}$, is merged with a subsonic structure calculated using the NLTE plane-parallel code TLUSTY (Hubeny \& Lanz 1995).\footnote{For our derived $T_{\ast}$ = 140 kK and $\log L/L_{\odot}$ = 3.6 we adopt $M = 0.6$ $M_{\odot}$ resulting in log $g$ = 6.1 for the TLUSTY model structure.} In order to reproduce the morphology of  He\,{\sc ii} $\lambda$4686 we obtain $v_{\infty}$ = 1900 km\,s$^{-1}$ and  $\beta \sim 2.5$, the former being similar to $v_\infty=2000$ km s$^{-1}$ as measured for the massive WN3 star WR152 (Crowther et al. 1995b). 
A clumped mass loss rate is calculated for an adopted volume filling factor of $f$ = 0.1, although this is poorly constrained for IC 4663. While $\beta$ values close to unity are normally found in weak-wind OB stars, larger values implying more extended accelerating zones are found in massive WR stars (e.g. Moffat 1996; L\'epine \& Moffat 1999).

Diagnostic optical lines of He\,{\sc ii} $\lambda$4686 and Ne\,{\sc vii} $\lambda$3890 enable the stellar temperature of $T_{\ast}$ = 140$\pm$20 kK to be obtained. Beyond this range, Ne\,{\sc vii} weakens at higher temperatures, where it is replaced by Ne\,{\sc viii} (e.g. $\lambda$4340) and at lower temperatures, where Ne\,{\sc vi} strengthens (e.g. $\lambda$4973). High lying recombination lines of N\,{\sc v} ($\lambda\lambda$4945) and O\,{\sc vi} ($\lambda$5290) serve as reliable N/He and O/He abundance diagnostics, from which mass fractions of 0.8\% and 0.05\% are obtained, respectively. Negligible hydrogen is required to reproduce the lower members of the Pickering-Balmer series, He\,{\sc ii} $\lambda$5412, He\,{\sc ii}+H$\beta$ and He\,{\sc ii} $\lambda$4542, permitting an upper limit of $\sim$2\% by mass. Since there are no suitable diagnostics for C/He we investigated whether we could place limits on the carbon mass fraction by recalculating the {\sc cmfgen} model with an elevated carbon abundance. The C~IV $\lambda$5801, 5812 emission lines became visible for values in excess of 0.1\%. 

The \textsc{cmfgen} model provides a satisfactory fit to the GMOS spectrum (Fig. \ref{fig:spec}), although O\,{\sc v} $\lambda$5590 is too weak, as is the blend of Ne\,{\sc viii} $\lambda$5668 and N\,{\sc v} $\lambda$5670. In addition, higher members of the He\,{\sc ii} Pickering series are predicted to exhibit weak P Cygni profiles, in contrast to observations. The theoretical spectral energy distribution (SED) was matched to the dereddened GMOS spectrophotometry enabling a determination of luminosity and mass loss rate. We obtained an interstellar extinction of $E(B-V)$ = $0.35\pm0.05$ ($A_{\rm V}$ = 1.1 mag) using a standard Galactic extinction law (Howarth 1983). This agrees well with the value of $E(B-V)$ = $0.4\pm0.1$ mag that Cavichia, Costa \& Maciel (2010, hereafter CCM10) measured from the Balmer decrement of the surrounding nebula. 

\begin{figure}
   \begin{center}
      \includegraphics[scale=0.35,angle=270]{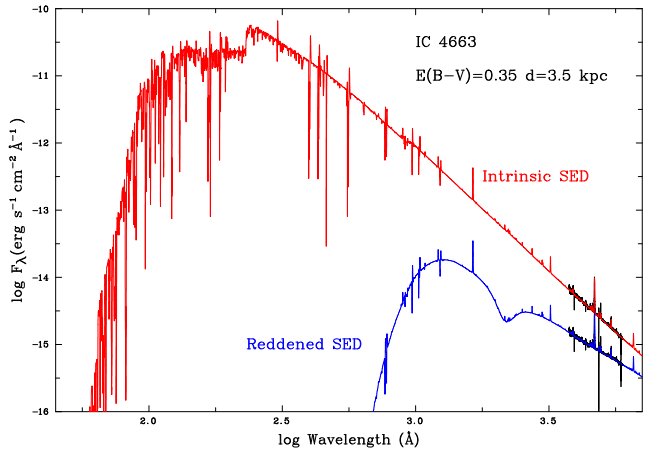}
   \end{center}
   \caption{Intrinsic and reddened SEDs of the NLTE model (red and blue, respectively) with the GMOS spectrum (black).}
   \label{fig:sed}
\end{figure}

We adopt a distance of 3.5 kpc from the statistical Shklovsky distance scale of Stanghellini, Shaw \& Villaver (2008, hereafter SSV08) which includes Magellanic Cloud PNe as calibrators. No individual error was given for the distance to IC~4663. We therefore take a conservative approach and assume a $2\sigma$ uncertainty of 60\%, which results in a distance of $3.5\pm2.1$ kpc. With this distance an absolute magnitude of $M_{\rm V}$ = +3.1 mag is derived for the [WN3] central star of IC~4663. The bolometric correction for our \textsc{cmfgen} model is $-$7.4, resulting in $M_{\rm Bol}$ = $-$4.3 or $\log L/L_{\odot}$ = 3.6. For this luminosity, a stellar radius of $R_{\ast}$ = 0.11 $R_{\odot}$ and a clumped mass loss rate of $\dot{M}$ = 1.8 $\times 10^{-8} M_{\odot}$\,yr$^{-1}$ are required. As indicated in Fig. \ref{fig:sed}, the SED of IC 4663 is exceptionally hard, with $Q_{0} = 10^{47.39}$ s$^{-1}$, $Q_{1} = 10^{47.27}$ s$^{-1}$, $Q_{2} = 10^{46.47}$ s$^{-1}$ for the H\,{\sc i}, He\,{\sc i} and He\,{\sc ii} continua, respectively. Table \ref{tab:stellar} summarises the stellar parameters of the [WN3] central star and includes ranges of distance-dependent quantities for the 2.1 kpc uncertainty in the distance.

\begin{table}
   \centering
   \caption{Stellar parameters of the [WN3] central star of IC~4663. Quantities in square brackets represent the range of parameter values for the 2.1 kpc uncertainty in distance.}
   \label{tab:stellar}
   \begin{tabular}{lll}
      \hline
      $T_\ast$ & $140\pm20$ & kK\\
      log $g$ & 6.1 & (assuming $M=0.6$ $M_\odot$)\\
      $v_\infty$ & 1900 & km s$^{-1}$\\
      $E(B-V)$ & $0.35\pm0.05$ & mag\\
      $V$ & $16.90\pm0.05$ & mag \\
      $d$ & $3.5\pm2.1$ & kpc (SSV08) \\
      $M_V$ & $3.1$ & mag [2.1--5.1]\\
      log $L/L_\odot$ & $3.6$ & [2.8--4.0]\\
      $R_\ast$ & $0.11$ & $R_\odot$ [0.04--0.17]\\
      $\dot{M}$ & $1.8$ & $10^{-8}M_{\odot}$\,yr$^{-1}$ [0.5--3.5]\\
      H & $<$2 & mass fraction (\%)\\
      He & 95 & mass fraction (\%)\\
      C & $<$0.1 & assumed mass fraction (\%)\\
      N & 0.8 &  mass fraction (\%)\\
      O & 0.05 & mass fraction (\%)\\
      Ne & 0.2 &  mass fraction (\%)\\
      \hline
   \end{tabular}
\end{table}

With the derived stellar properties of IC~4663 we can quantitatively judge its wind strength against massive WN3 stars. Table \ref{tab:rt} summarises the stellar properties of massive WN3 stars (Hamann et al. 2006) and IC~4663, together with the transformed radius $R_t$ (see Eqn. 1 of Hamann et al. 2006). The magnitude of $R_t$ diagnoses the wind density (i.e. the emission-line strength) which is essentially the same for massive WR and [WR] stars of a given spectral type, despite the order of magnitude difference in radii (e.g. Crowther et al. 2006). Low values of $R_t$ correspond to strong, thick winds and high values to weak, thin winds. Since the $R_t$ of IC~4663 lies between WR152 and WR3, the stellar wind is similarly dense to massive WN3 stars, entirely consistent with our morphological [WN3] classification. This argues against any weak wind classification of the central star of IC~4663 (e.g. Of or Of-WR, see M\'endez, Herrero \& Manchado (1990) and M\'endez (1991)).

\begin{table}
   \centering
   \caption{Stellar properties and transformed radii $R_t$ for three massive WN3 stars (Hamann et al. 2006) and the [WN3] central star of IC~4663. The $D$ quantity is the wind clumping factor.}
   \label{tab:rt}
   \begin{tabular}{lrrrrrr}
      \hline
      Object & $R_*/R_\odot$ & log $L/L_\odot$ &$v_\infty$ & log $\dot{M}$ & $D$ & $R_t/R_\odot$\\
      \hline
      WR3     & 2.65 & 5.6& 2700 & -5.3  & 4 & 15.1\\
      WR46    & 2.11 & 5.8& 2300 & -5.1 & 4 & 6.8\\
      WR152   & 2.20 & 5.3& 2000 & -5.5  & 4 & 11.9\\
\noalign{\smallskip}     
      IC~4663 & 0.11 &3.6 & 1900 & -7.7& 10 & 13.4\\
      \hline
   \end{tabular}
\end{table}

\subsection{Variability}
\label{sec:var}
\begin{figure}
      \begin{center}
         \includegraphics[scale=0.35,angle=270]{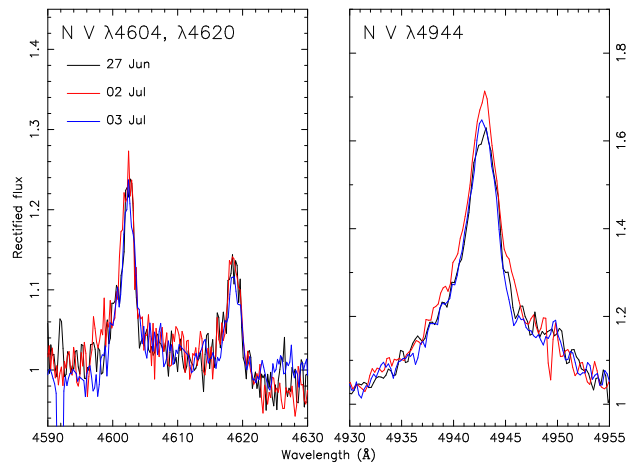}
      \end{center}
      \caption{Line profiles of N~V $\lambda$4604, $\lambda$4620 and $\lambda$4944 in the overlap region between all three epochs. Note the absence of any RV shifts $\ga$5--10 km s$^{-1}$ that argues against the presence of a post common-envelope binary companion.}
      \label{fig:rv}
   \end{figure}

The GMOS observations were taken on three separate epochs, namely $E1$ (June 27), $E2$ (July 2) and $E3$ (July 3), separated by $E2-E1=5.273$ d, $E3-E2=1.005$ d and $E3-E1=6.278$ d. Figure \ref{fig:rv} depicts the narrow N~V emission lines which show a root mean square radial velocity (RV) variability of 3--4 km s$^{-1}$ using cross-correlation techniques, i.e. no evidence of variability $\ga$5--10 km s$^{-1}$. A main-sequence or white dwarf companion in a post common-envelope binary (i.e. $P\la1$ day) would produce larger variability than our detection limit and is therefore unlikely to be present unless the orbital inclination is very low. The lack of temporal coverage less than a day also makes this possibility difficult to rule out entirely since the median period of close binary central stars is 0.3--0.4 days (Miszalski et al. 2009a, 2011b). The small variability in the line intensity profile may be explained by either noise in the spectrum or wind turbulence (e.g. Grosdidier et al. 2000, 2001b). There is little historical information on any photometric variability of IC~4663, e.g. whether an LMC-N66-like outburst has occurred. The insensitivity of Shaw \& Kaler (1989) to magnitudes fainter than $V=16$ provides a reasonable lower limit to the central star magnitude in 1989. The very bright nebula precludes measuring central star magnitudes from photographic plate material.

\section{The nebula}
\label{sec:nebula}
\subsection{Observations and basic properties}
\label{sec:nebobs}
  The nebulae surrounding WR central stars contain valuable additional information to better understand the nature of their nuclei. Acquisition images of IC~4663 were taken during our GMOS programme through the filters Ha (120 s), OIII (120 s) and OIIIC (90 s) whose central wavelengths/FWHMs are 656.0/7.2 nm, 499.0/4.5 nm and 514.0/8.8 nm, respectively. The seeing was 1.5\arcsec\ (FWHM) on 27 June and 1.1\arcsec\ (FWHM) on 2 and 3 July as measured from the OIIIC images. 
 
\begin{figure*}
     \begin{center}
       \includegraphics[scale=0.60]{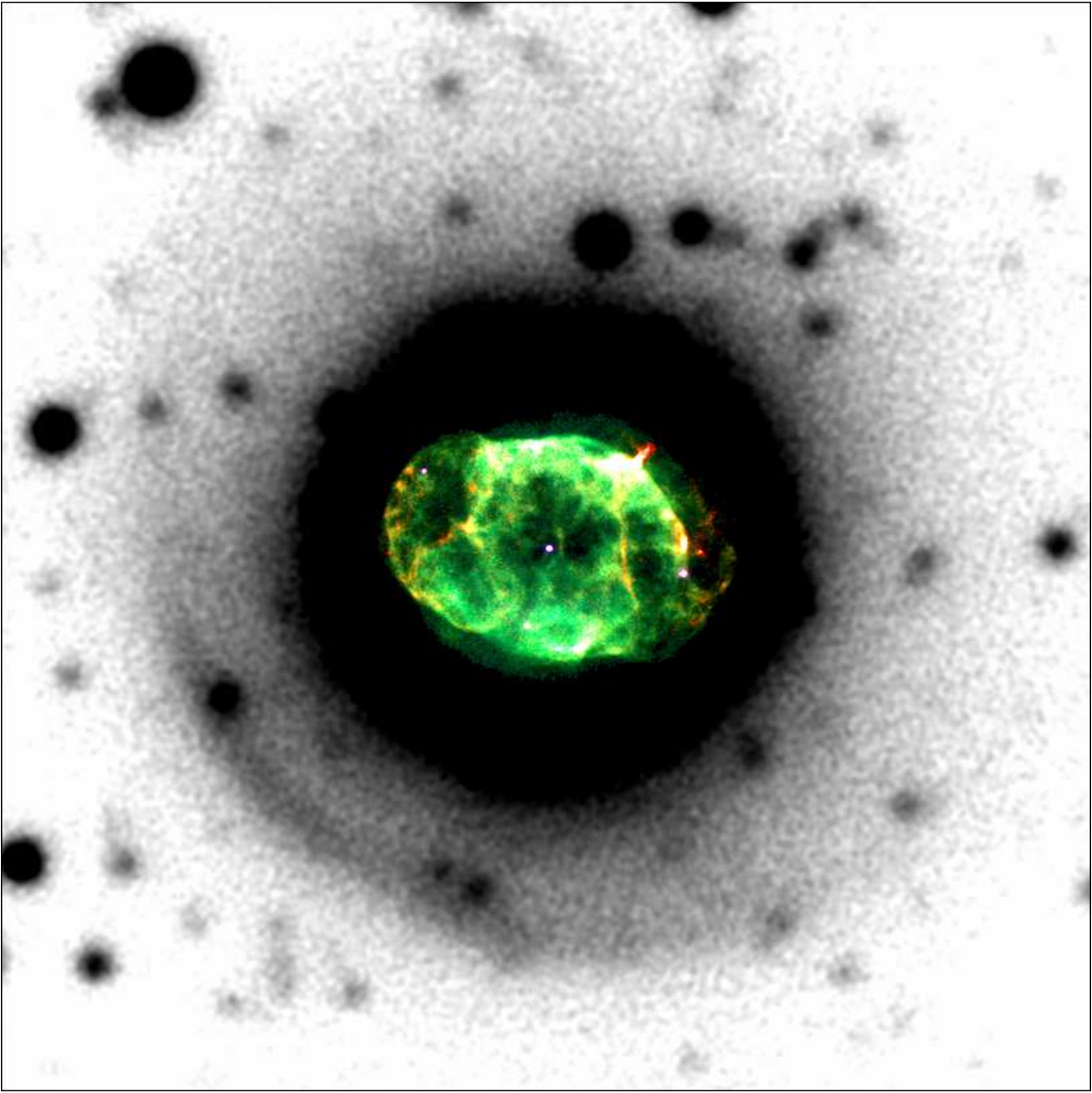}
     \end{center}
     \caption{Gemini South GMOS [O~III] image of IC~4663 (background) showing the newly discovered AGB halo, overlaid with an \emph{HST} colour-composite image (red, green and blue channels were made from $F658N$, $F502N$ and $F555W$ images, respectively). The image is centered on the central star and measures $60\times60$ arcsec$^2$ ($\sim$1 pc across at 3.5 kpc) with North up and East to left.}
     \label{fig:oiii}
  \end{figure*}

  Figure \ref{fig:oiii} shows the GMOS [O~III] image overlaid with a colour-composite image made from the \emph{HST} images published by Hajian et al. (2007). Revealed for the first time is a low surface brightness AGB halo (Corradi et al. 2003). A brightened rim lies 22\arcsec\ SE of the central star and is likely caused by ISM interaction. At $\sim$0.25\% of the main nebula peak [O~III] intensity, i.e. 400 times fainter, roughly corresponding to the extent visible in the scaling selected for Fig. \ref{fig:oiii}, the halo measures 52.5\arcsec\ $\times$ 55.0\arcsec\ across. The halo may even reach as far as 72\arcsec\ across before it becomes indistinguishable from the sky background. 
  
  The PN morphology is best captured by the aforementioned Hajian et al. (2007) \emph{HST} images that were taken on 04 August 1998 through the $F502N$ ([O~III]), $F658N$ ([N~II]) and $F555W$ filters. The data were retrieved from the HLA and have total exposure times of 520 s, 320 s and 1020 s, respectively. A nebula diameter of 16.0\arcsec $\times$ 19.5\arcsec\ was measured at 10\% of peak intensity in the $F502N$ image. The overall morphology appears to be an ellipsoidal bubble surrounded by a second fainter shell similar to other bright PNe (e.g. NGC~1501). This interpretation is consistent with the high resolution spectrograms published by Hajian et al. (2007). As previously mentioned, low-ionisation filaments are also present (Gon{\c c}alves et al. 2001, visible as red filaments in Fig. \ref{fig:oiii}) and were part of the selection criteria for the inclusion of this PN in our GMOS programme. 
 
  Combining the equatorial expansion velocity of 30 km s$^{-1}$ measured by Hajian et al. (2007), with the minor axis nebula diameter of 16\arcsec\ and our assumed distance of 3.5 kpc (SSV08), gives an expansion age of $t_\mathrm{neb}\sim4400$ years. For the halo we assume a 15 km s$^{-1}$ expansion velocity (Corradi et al. 2003) to find $t_\mathrm{halo}\sim30500$ years and $t_\mathrm{halo}-t_\mathrm{neb}\sim26000$ years.  We derive a heliocentric RV $v_\mathrm{hrv}$ of $-78\pm3$ km s$^{-1}$ from the brightest emission lines in the GMOS nebula spectra. This value is in agreement with $-74$ km s$^{-1}$ measured by Beaulieu et al. (1999). Shaw \& Kaler (1989) measured integrated fluxes for the nebula and found $\log F(H\beta) =-11.44$ ($-10.93$ dereddened), where $F(H\beta)$ is in units of erg cm$^{-2}$ s$^{-1}$. Table \ref{tab:basic} summarises the basic observed properties of IC~4663.

   \begin{table}
      \centering
      \caption{Basic nebula properties of IC~4663.}
      \label{tab:basic}
      \begin{tabular}{lrl}
         \hline
         PN G & 346.2$-$08.2 &\\
         RA & 17$^\mathrm{h}$45$^\mathrm{m}$28\fs6 & \\
         Dec. & $-$44$^\circ$54$'$17 &\\
         Nebula diameter & 16.0\arcsec\ $\times$ 19.5\arcsec & this work\\
         Halo dimensions & 52.5\arcsec\ $\times$ 55.0\arcsec & this work\\
         $\log F(H\beta)$ & $-$11.44 & Shaw \& Kaler (1989)\\
         $v_\mathrm{exp}$ & 30 km s$^{-1}$ & Hajian et al. (2007)\\
         $v_\mathrm{hrv}$ & $-78\pm3$ km s$^{-1}$ & this work\\
         $t_\mathrm{neb}$ & $\sim$4400 yrs & this work\\
         $t_\mathrm{halo}$ & $\sim$30500 yrs & this work\\
         \hline
      \end{tabular}
   \end{table}

\subsection{Chemical abundances and plasma parameters}
\label{sec:plasma}
\begin{figure*}
   \begin{center}
      \includegraphics[scale=0.8,angle=270]{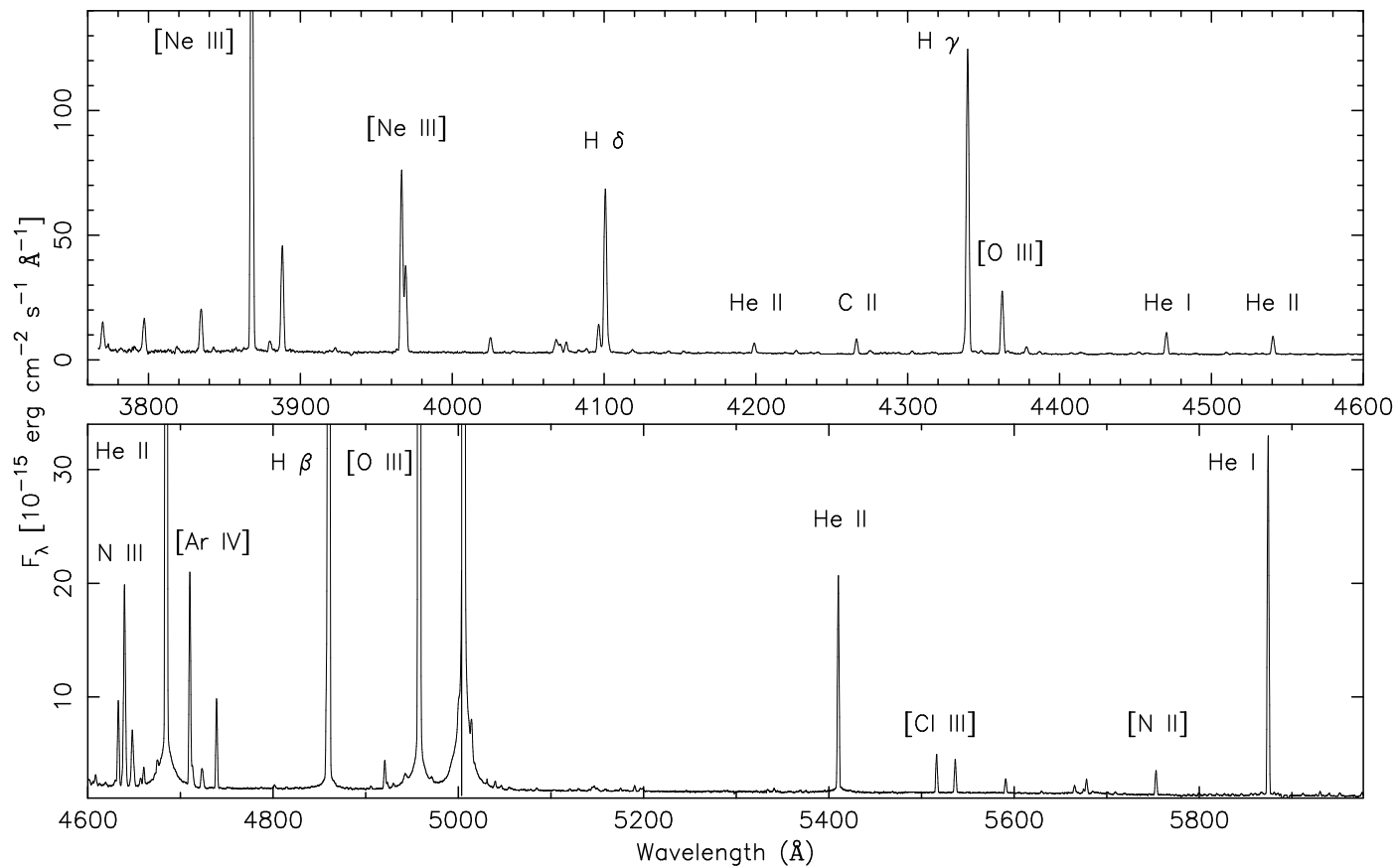}
   \end{center}
   \caption{GMOS nebular spectrum of IC~4663 with the main lines identified. See Tab. \ref{tab:optical} for line intensities.}
   \label{fig:neb}
\end{figure*}

The nebular spectrum of IC~4663 was previously analysed by CCM10 using low resolution ($\sim$6\AA\ FWHM) spectroscopy obtained with a 1.6-m telescope. Figure \ref{fig:neb} shows our combined nebular spectrum made from averaging both of the 2400 s GMOS exposures. We compared the measured emission line intensities in our deep GMOS spectrum with the observations of CCM10 and found a good correlation with the line fluxes for the stronger lines, but also a tendency for the weak line fluxes to be overestimated by CCM10. For instance, the He I lines at $\lambda$4471 and $\lambda$6678 are twice as bright as expected from the $\lambda$5876 line, whereas all four lines in our spectrum are close to their theoretical ratios. Similarly, the fluxes of [Ar~IV] $\lambda$7170 and $\lambda$7237 are $\sim$3 times stronger than predicted from their $\lambda$4740 line. Thus we shall ignore them in our analysis. Our [Ar~IV] $\lambda$4740 line is only 1/3 of the CCM10 value.

Since the GMOS spectrum is deeper with greater S/N we produced a composite spectrum by using line intensities from our deep GMOS spectrum for the blue part ($\lambda\le5900$\AA) and CCM10 for the red part (except for [O~II] $\lambda$3728 which lies outside our GMOS spectrum). The published intensities of CCM10 were reddened with the reddening law of Fitzpatrick (1999) for $R_V = 3.1$, and the fluxes were scaled to the GMOS H$\beta$ flux. We measured UV line fluxes from archival \emph{IUE} spectra SWP33946 and LWP13707, observed on 20 July 1988, that were joined to the optical spectrum, taking into account the finite coverage of the \emph{IUE} aperture. Although the aperture includes the central star, the low-dispersion \emph{IUE} spectra only show nebula emission lines and no features from the central star. We estimate the stellar contribution to the \emph{IUE} aperture at He~II $\lambda$1640 to be $\sim$5\% in comparison with our {\sc cmfgen} model.

Tables \ref{tab:iue} and \ref{tab:optical} contain the full composite nebula spectrum of IC~4663. The line fluxes were analysed with the plasma diagnostics program \textsc{hoppla} (Acker et al. 1991; K\"oppen et al. 1991; Girard et al. 2007) which first determines the extinction constant - using the $R_V = 3.1$ reddening law of Fitzpatrick (1999) - as well as the electron temperature and density in a consistent way. The H$\alpha$/H$\beta$ ratio would give $c=0.58$, the inclusion of the blue Balmer lines gives a slightly smaller value of $0.54$. We adopted $c=0.51$ to be consistent with the dereddening of the stellar spectrum. In either case, all Balmer lies are fit within 10\% of their intensity from case B recombination theory. The He~I line intensities are matched better than 10\% by case B recombination emissivities corrected for collisional excitation (Porter, Ferland \& MacAdam 2007), except for the $\lambda$6678 and $\lambda$7065 lines from CCM10. We note that the choice of extinction constant has only very little influence on the derived elemental abundances (less than 0.04 dex). 

With the adopted extinction, we then determined the aperture factor for the \emph{IUE} spectrum by minimizing the error 
in the He~II UV and optical lines $\lambda\lambda$1640, 4686, 3203, according to case B recombination theory. This scaling factor is also consistent with the ratio of the nebular area and the \emph{IUE} aperture (an ellipse measuring 10\arcsec$\times$20\arcsec). 

\begin{table}
     \centering
     \caption{UV emission line intensities measured from the \emph{IUE} spectra of IC~4633 (SWP33946 and LWP13707) scaled to H$\beta = 100$. Since the \emph{IUE} aperture does not cover the entire nebula, the fluxes were scaled as to give the best agreement between the intensities of the UV and optical He~II lines. The measured fluxes in units of $10^{-13}$ erg cm$^{-2}$ s$^{-1}$ \AA$^{-1}$ may be obtained by dividing the flux values by 5.5.}
        \label{tab:iue}
     \begin{tabular}{lrr}
        \hline
        Line & Flux & Intensity\\ 
        \hline
CIV 1550    &   $<8$   &   -- \\  
HeII 1640   &   124  &  558 \\   
{}[NIII] 1753 &  6:   &   27: \\  
{}CIII] 1908 & 36   &  179 \\    
{}[NeIV] 2424 & 25   &  106 \\   
HeII 2734   &   10   &   28 \\   
OIII 2837*  &    8:   &   20: \\  
OIII 3047*  &   11   &   26 \\   
OIII 3133*  &   76   &  162 \\   
HeII 3204   &   20   &   41 \\   
OIII 3299*  &   18  &   36 \\    
OIII 3341*  &   37   &   71 \\   
    \hline
  \end{tabular}
  \begin{flushleft}
     $^*$ Bowen fluorescence line.
  \end{flushleft}
\end{table}

  \begin{table}
     \centering
     \caption{Nebular emission line intensities of IC~4663 scaled to H$\beta = 100$ as measured from our GMOS spectrum (first half) and CCM10 (second half). The CCM10 fluxes were obtained by reddening the published intensities with $c=0.58$ and the $R_V=3.1$ law of Fitzpatrick (1999).} 
        \label{tab:optical}
     \begin{tabular}{lrr}
        \hline
           Line  & Flux   & Intensity  \\
        \hline
HI 3798       &    4.8   &    6.6 \\
HI 3835       &    6.2   &    8.5 \\
{}[NeIII] 3869  &   79.4   &  106.8 \\
HeI 3889      &   15.4   &   20.6 \\
HeII 3924     &    0.7   &    0.9 \\
{}[NeIII]+HI 3967 & 24.6   &   32.1 \\
HI 3970       &   11.5   &   15.0 \\
HeI 4026      &    2.3   &    3.0 \\
{}[SII] 4068    &    2.8   &    3.6 \\
OII 4072      &    1.5   &    1.9 \\
{}[SII] 4076    &    1.5   &    1.9 \\
NIII 4097     &    4.2   &    5.2 \\
H$\delta$ 4102       &   24.2   &   30.5 \\
OII 4119      &    0.7   &    0.8 \\
HeII 4200     &    1.4   &    1.7 \\
CII 4267      &    2.1   &    2.5 \\
H$\gamma$ 4340      &   43.9   &   51.5 \\
{}[OIII] 4363   &    8.8   &   10.3 \\
NIII 4379     &    1.2   &    1.3 \\
HeI 4388      &    0.45  &    0.4 \\
HeI 4472      &    3.1   &    3.5 \\
NIII 4511     &    2.5   &    2.8 \\
NIII 4634     &    2.8   &    3.0 \\
NIII 4641     &    7.3   &    7.8 \\
OII 4649      &    2.5   &    2.6 \\
HeII 4686     &   80.6   &   85.3 \\
{}[ArIV]+HeI 4711 &  6.9   &    7.0 \\
{}[NeIV] 4714   &    0.8   &    0.8 \\
{}[NeIV] 4724   &    0.9   &    0.9 \\
{}[ArIV] 4740   &    2.7   &    2.8 \\
H$\beta$ 4861       &    100.0   &    100.0 \\
HeI 4922      &    0.9   &    0.9 \\
{}[OIII] 4959   &  375.4   &    364.0 \\
HeII 5412     &    6.5   &    5.5 \\
{}[ClIII] 5518  &    1.2   &    1.0 \\
{}[ClIII] 5538  &    1.0   &    0.8 \\
OIII 5592     &    0.4  &    0.4 \\
NII 5667      &    0.3   &    0.2 \\
NII 5680      &    0.5   &    0.4 \\
{}[NII] 5754    &    0.8   &    0.6 \\
HeI 5876      &   11.8   &    9.0 \\
     \hline
{}[OII] 3728    &   38.9   &   54.6 \\
{}[OIII] 5007   & 1085.0   & 1035.1 \\
{}[SIII] 6312   &    3.9   &    2.7 \\
{}[ArV] 6435    &    0.8   &    0.5 \\
{}[NII] 6548    &   31.3   &   20.8 \\
H$\alpha$ 6563       &  454.2   &  301.5 \\
{}[NII] 6583    &   50.1   &   33.5 \\
HeI 6678      &    5.2   &    3.4 \\
{}[SII] 6716    &    3.9   &    2.6 \\
{}[SII] 6731    &    4.9   &    3.2 \\
{}[ArV] 7006    &    1.9   &    1.2 \\
HeI 7065      &    2.5   &    1.5 \\
{}[ArIII] 7136  &   50.9   &   30.8 \\
{}[ArIV] 7170   &    2.1   &    1.3 \\
{}[ArIV] 7237   &    2.0   &    1.1 \\
{}[OII] 7325    &    5.1   &    3.0 \\
{}[ArIII] 7751  &   11.9   &    6.6 \\
    \hline
  \end{tabular}
\end{table}

Ionic abundances were derived from each of the dereddened line intensities, listed in Tab. \ref{tab:ionic}, using the electron temperatures for the low, middle, and high ionic species
(given in Tab. \ref{tab:plasma}), and weighted with the emissivity of each line. Elemental abundances were obtained by summing over the ionic abundances and correcting for unobserved stages of ionisation:

\[
     \rm {He\over H} = {He^+\over H^+} + {He^{++}\over H^+}
\]
Carbon is derived solely from the C~III] emission line (see below):
\[
     \rm {C\over H} = {C^{++}\over H^+}
\]
We use the standard correction factors (e.g. Aller 1984):

\[
\rm {N\over H} = {N^+\over H^+} {O/H\over O^+/H^+}
\]

\[
     \rm {O\over H} = \left({O^+\over H^+} + {O^{++}\over H^+}\right) {He/H \over He^{++}/H^+}
\]
For neon, we use

\[
     \rm {Ne\over H} = {Ne^{++}\over H^+}  {O/H \over O^{++}/H^+}
\]
and note that the sum of Ne$^{++}$ and Ne$^{3+}$ gives only a slightly higher 
abundance of $12+\log({\rm Ne/H}) = 8.17$. 
For sulphur and chlorine, we employ the ionisation correction factors derived from photoionisation models (K\"oppen, Acker \& Stenholm 1991):

\[
     \rm {S\over H} = \left({S^+\over H^+} + {S^{++}\over H^+}\right) 
      \left[1.43+0.196 \left({O^{++}\over O^{+}}\right)^{1.29}\right]
\]

\[
     \rm {Cl\over H} = {Cl^{++}\over H^+}  \left({He/H \over He^{++}/H^+}\right)^2
\]

Argon is treated by the correction from Barker (1980):

\[
     \rm {Ar\over H} = \left({Ar^{++}\over H^+} + {Ar^{3+}\over H^+} 
                       + {Ar^{4+}\over H^+}\right)  {S^{+} + S^{++}\over S^{++}}
\]

The resulting abundances are collated in Tab. \ref{tab:plasma}. 
A few remarks are necessary. 
Our helium abundance is slightly higher than found by CCM10. This is due
to the intensity of the He~I $\lambda$5876 line being larger --
by a factor of 1.4 -- in our GMOS spectrum. Due to the ionisation
corrections this also results in lower oxygen and neon abundances. 
The intensity ratio of [N~II] 6583/6548 (from CCM10) of 1.6 is substantially lower than the theoretical ratio of 2.9, which indicates problems with the separation of the lines from H$\alpha$. As to make use of all information, we use both lines, and we use the $\lambda$5755 line from the GMOS spectrum. These choices influence our results on nitrogen, albeit
not in a critical way. Taking the $\lambda$6583 line only increases
the electron temperature in the O$^+$ zone to 11200 K, giving
a slightly lower nitrogen abundance of 8.22, and N/O$ = -0.48$.
If we took also the $\lambda$5755 intensity from CCM10 -- which
is larger by a factor of 1.5 than the GMOS value -- the electron
temperature increases further to 13900 K, the nitrogen abundance
to 8.27, and N/O$ = -0.35$, the same value as derived by CCM10.
Furthermore, from the just discernible [N~III] $\lambda$1753 line
in the IUE spectrum, we may derive an ionic abundance which is
substantially larger than from N$^+$. Since in a high excitation
nebula nitrogen is found predominantly in the form of N$^{++}$,
the derived nitrogen abundance of $12+\log({\rm N/H})=8.11$ is
a little less than that obtained from the [N~II] lines, but
with a substantial ionisation correction factor (30--60 in
the various cases considered here). This gives N/O=$-0.59$.
We conclude that the nitrogen abundance is certainly in the range 8.1--8.3, 
hence N/O = $-0.6$ to $-0.3$. This is just below the threshold value for a bona-fide Type I PN in Peimbert's classification and well below the (linear) value of 0.8 for a Type I in the sense of Kingsburgh \& Barlow (1994).

 The C$^{++}$ abundance derived from the optical recombination line at 4267 \AA\ is about 30 times higher than from the collisionally excited doublet C~III] 1908 \AA . This abundance discrepancy factor is larger than the usual factor of 2--3, but other extreme examples are known (e.g. Liu et al. 2006). The presence of N~III and O~III recombinations lines is in agreement with the very high level of nebular excitation, which is in concordance with the UV flux distribution predicted from our model atmosphere for the central star. A recombination line analysis of the nebula spectrum will not be performed here.

  \begin{table}
     \centering
     \caption{Ionic abundances (relative to H$^+$) and the emission 
        lines from which they were derived.} 
        \label{tab:ionic}
     \begin{tabular}{lll}
     \noalign{\smallskip} \hline  \noalign{\smallskip}
         Ion & Ionic abundance & Lines used (\AA)  \\
     \noalign{\smallskip} \hline
     \noalign{\smallskip}
      He$^+$    & $6.43\times10^{-2}$  & 5876, 4471, 6678, 7065, 4922, 4388 \\
      He$^{++}$ & $6.58\times10^{-2}$  & 1640, 4686, 3203, 2734, 5411, 4199, 3923 \\
      C$^{++}$  & $1.57\times10^{-4}$  & 1908 \\
      N$^+$     & $6.42\times10^{-6}$  & 6583, 6548, 5755 \\
      N$^{++}$  & $1.22\times10^{-4}$  & 1753 \\
      O$^+$     & $1.79\times10^{-5}$  & 3728, 7325 \\
      O$^{++}$  & $2.32\times10^{-4}$  & 5007, 4959, 4363 \\
      Ne$^{++}$ & $5.92\times10^{-5}$  & 3869 \\
      Ne$^{3+}$ & $8.73\times10^{-5}$  & 2424, 4714, 4724 \\
      S$^+$     & $5.02\times10^{-7}$  & 4076, 4068, 6717, 6731 \\
      S$^{++}$  & $4.78\times10^{-6}$  & 6312 \\
      Cl$^{++}$ & $7.66\times10^{-8}$  & 5517, 5537 \\
      Ar$^{++}$ & $1.96\times10^{-6}$  & 7135, 7750 \\
      Ar$^{3+}$ & $3.38\times10^{-7}$  & 4740 \\
      Ar$^{4+}$ & $1.42\times10^{-7}$  & 7005, 6434 \\
    \noalign{\smallskip} \hline
  \end{tabular}
\end{table}

  \begin{table}\centering
     \caption{Plasma diagnostics and the chemical composition of
        IC~4663, given in the usual notation 
        $12+\log[n({\rm X})/n({\rm H})]$, from this work and from CCM10. 
        For comparison the average abundances for Type I and non-Type I
        PNe are taken from Kingsburgh \& Barlow (1994) and 
        Solar abundances from Asplund et al. (2009).} 
        \label{tab:plasma}
     \begin{tabular}{lrrrrrr}
     \noalign{\smallskip} \hline  \noalign{\smallskip}
     & \multicolumn{2}{c}{IC~4663} &  Type I & non- & Sun \\
                         & this work  &  CCM10           &        & Type I & \\
     \noalign{\smallskip} \hline
     \noalign{\smallskip}
      $c$                & 0.51 & 0.58  & --- & --- & --- \\
      $n_e$ [cm$^{-3}$]   &  1000  & 3810  & --- & --- & --- \\
      $T_e(\rm{O}^+)$ [K]     & 10300  & 12800 & --- & --- & --- \\
      $T_e(\rm{O}^{++})$ [K]  & 11500  & 11500 & --- & --- & --- \\
      $T_e(\rm{He}^{++})$ [K] & 11400  &  ---  & --- & --- & --- \\
     \noalign{\smallskip} \hline
     \noalign{\smallskip}
      He    & 11.11 & 11.07 & 11.11 & 11.05 & 10.93 \\
      C     &  8.20 &  ---  & 8.43 & 8.81   & 8.43  \\
      N     &  8.26 &  8.49 & 8.72  & 8.14  & 7.83  \\
      O     &  8.70 &  8.84 & 8.65  & 8.69  & 8.69  \\
      Ne    &  8.11 &  8.25 & 8.09  & 8.10  & 7.93  \\
      S     &  7.02 &  7.03 & 6.91  & 6.91  & 7.12  \\
      Cl    &  5.50 &  ---  & ---   & ---   & 5.50  \\
      Ar    &  6.43 &  6.87 & 6.42  & 6.38  & 6.40  \\
     \noalign{\smallskip} \hline
     \noalign{\smallskip}
     log(C/O)  & $-0.50 $  & --- & $-0.12$ & $+0.12$ & $-0.26$ \\
      log(N/O)  & $-0.45 $  & $-0.35$ & $+0.07$ & $-0.55$ & $-0.86$ \\
      log(Ne/O) & $-0.59 $  & $-0.59$ & $-0.56$ & $-0.59$ & $-0.76$ \\
      log(S/O)  & $-1.68 $  & $-1.81$ & $-1.74$ & $-1.78$ & $-1.57$ \\
      log(Cl/O) & $-3.20 $  & ---     & ---      & ---    & $-3.19$ \\
      log(Ar/O) & $-2.27 $  & $-1.97$ & $-2.23$ & $-2.31$ & $-2.29$ \\
    \noalign{\smallskip} \hline
  \end{tabular}
\end{table}

 Despite the high nebular excitation the \emph{IUE} spectrum shows no detectable
 C~IV $\lambda$1550 emission. 
 In nebulae with a large central hole, this line is strongly 
 weakened, as the C$^{3+}$ zone is remote from the central star, and hence 
 the ionizing radiation is geometrically strongly diluted (c.f. K\"oppen 1983).
 The monochromatic images of IC~4663 indeed suggest a bubble-type geometry for
 the emission region. Tests with photoionisation models indicate that in a 
 thin-shell type nebula the C~IV lines can be suppressed sufficiently to agree with the \emph{IUE} spectrum. As a consequence, we may assume that the C$^{++}$ zone
 nearly completely fills the nebula, and that the carbon abundance derived 
 from the C~III] 1908 lines may be close to the true value. Since the nebula 
 has a rather elliptical shape and also shows indications for a double shell,
 we refrain from constructing a spherical photoionisation model.

The derived abundances show that our He/H is slightly higher than in 
CCM10 and closer to a Type~I composition (Peimbert \& Torres-Peimbert 1983; Kingsburgh \& Barlow 1994), because the He~I 5876 intensity from the GMOS
spectrum is higher. We note that our value is also consistent with the 
other He~I lines. However, the N/O ratio is not as high as to classify 
it as a genuine Type~I nebula. 
The nearly Solar oxygen abundance is also similar to the average PN,
as are the Ne/O and S/O ratios. The ratios of the $\alpha$-elements
are close to the average values found in H~II regions of spiral and 
irregular galaxies (Henry \& Worthey 1999), namely Ne/O $= -0.67$ (slightly 
lower than in PN and in IC~4663), S/O $= -1.55$ (slightly higher than in IC~4663), and Ar/O $= -2.25$. Since these elements are not expected to be affected by nucleosynthesis in the progenitor star, their ratios reflect the interstellar medium values which are determined by nucleosynthesis in massive stars.

The carbon to oxygen ratio is lower than in both PN types and the Sun, but in light of both 
the strong sensitivity of the UV lines to electron temperature and the 
uncertainty of the ionisation correction factor for carbon, this difference 
of 0.3 dex cannot be taken as a solid proof for a carbon underabundance.
If placed in Fig. 7 of Henry \& Worthey (1999), IC~4663 would be located near the lower envelope of scattered points, but would still be consistent with the average relation. 
Finally, the chlorine to oxygen ratio is Solar. 
The abundance pattern of IC~4663 is close to the Sun, except for enhancements in helium, nitrogen and neon.

  \subsection{Evolutionary status}
  \label{sec:nebevol}
\begin{figure}
   \begin{center}
      \includegraphics[scale=0.35,angle=270]{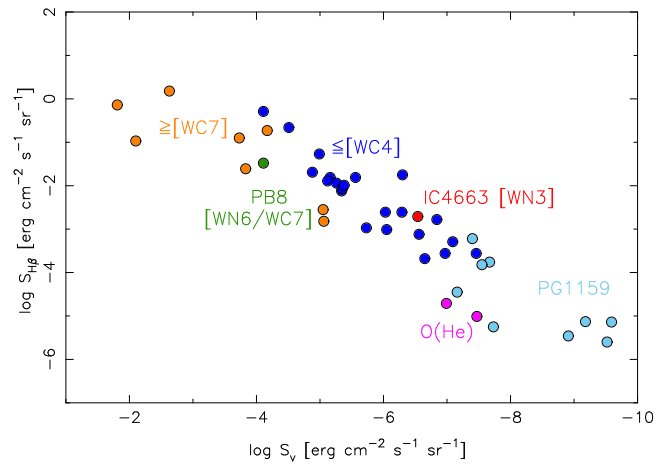}
   \end{center}
   \caption{Hydrogen-deficient PNe in the distance-independent log $S_V$ and log $S_{H\beta}$ plane (see G\'orny \& Tylenda 2000). }
   \label{fig:gorny}
\end{figure}

 The evolutionary state of the PN IC~4663 may also be compared against other PNe using the distance-independent parameters log $S_V$ and log $S_{H\beta}$ (G\'orny \& Tylenda 2000). The parameters are defined as the H$\beta$ surface brightness and an analagous quantity replacing $F_{H\beta}$ with $F_V$, an extinction-corrected stellar flux in the $V$-band, respectively. Both $S_V$ and $S_{H\beta}$ decrease as a PN expands and evolves towards the WD state. Figure \ref{fig:gorny} portrays the position of IC~4663 amongst other PNe with H-deficient central stars (log $S_V=-6.54$ and log $S_{H\beta}=-2.71$) and PB~8 (log $S_V=-4.11$ and log $S_{H\beta}=-1.48$, based on the updated parameters in Todt et al. 2010a). The location of IC~4663 is coincident with PNe that have a [WO] central star (i.e. [WC4] or earlier) and PB~8 with those that have a [WC] central star (i.e. [WC7] or later). 

  \section{Discussion}
  \label{sec:discussion}

   \subsection{A bona-fide PN}
   \label{sec:bona}
   Table \ref{tab:dist} delineates three illustrative distance scenarios for IC~4663. At an assumed distance of 3.5 kpc (SSV08), the central star has $M_V=+3.1$ mag. This is over 6 magnitudes fainter than $M_V=-3.0$ mag, the typical luminosity for a massive WN3 star (Crowther et al. 1995b; Hamann et al. 2006). The other two scenarios demonstrate that the sub-luminous nature of the central star is robust against any reasonable uncertainties that may be associated with the distance of SSV08. Even if IC~4663 were a member of the Galactic Bulge at $d=8.0$ kpc, then the central star is still at least 4 magnitudes too faint. IC~4663 cannot however be a member of the Bulge because its angular diameter and H$\beta$ flux are anomalously large compared to most Bulge PNe (see Fig. 7 of G\'orny et al. 2009). Indeed, an implausible distance of 58 kpc would be required to reproduce $M_V=-3.0$ mag, therefore for any reasonable distance estimate the central star must be sub-luminous and belong to the PN. At 3.5 kpc IC~4663 would lie 500 pc below the Galactic Plane, i.e. between the bias-corrected scale heights of the thin (300-375 pc) and thick (900-940 pc) disks (Siegel et al. 2002; Juri\'c et al. 2008). Massive WR stars that are not runaway stars would be restricted to the young disk population which has a scale height of only $\sim48\pm3$ pc (Bonatto et al. 2006). This intermediate height might on its own suggest a thick disk progenitor, however the solar oxygen and argon abundances, which reflect the metallicity of the gas which formed the progenitor star, suggests IC~4663 is a thin disk object.

         \begin{table}
            \centering
            \caption{Absolute visual magnitudes and vertical displacement from Galactic Plane for the central star of IC~4663 at a range of distances ($V=16.90$ mag and $E(B-V)=0.35$ mag, Sect. \ref{sec:nebobs}).}
            \label{tab:dist}
            \begin{tabular}{rccl}
               \hline
                 $d$ & $M_V$ & $z$ & Comment\\
                 (kpc)     &   & (kpc) & \\
                 \hline
               3.5  & $+$3.1 & $-$0.50 & $d$ from SSV08\\
               8.0  & $+$1.3 & $-$1.14 & $d$ of Galactic Bulge PN\\
               58.0 & $-$3.0 & $-$8.27 & $M_V$ of massive WN3 star\\
               \hline
            \end{tabular}
         \end{table}

      The presence of an AGB halo is a telltale signature of a PN (Corradi et al. 2003). The ellipsoidal morphology of the main nebula, its relatively low expansion velocity and density are all typical of PNe. Its evolutionary status overlaps with other PNe in Fig. \ref{fig:gorny} and exhibits a highly ionised emission line spectrum with a typical chemical abundance pattern for a PN.

  \subsection{A second post-AGB H-deficient and He-rich evolutionary sequence}
  Table \ref{tab:comp} contrasts the He-dominated composition of the [WN3] central star of IC~4663 against those of other H-deficient post-AGB stars.\footnote{Some caution should be taken when comparing compositions derived from different analysis techniques (see e.g. Werner \& Herwig 2006).} Clearly IC~4663 does not fit the carbon- and oxygen-rich evolutionary sequence [WCL]$\to$[WCE]$\to$PG1159 (Werner \& Herwig 2006). Rather, it fits amongst other rare He-dominated objects including RCB stars, O(He) stars, extreme helium B stars and helium-rich subdwarf O stars for which there is no accepted explanation for their origin (Werner \& Herwig 2006). The born-again evolutionary models simply cannot reproduce the helium-dominated atmospheres (e.g. Herwig 2001). Some proposed explanations include a double-degenerate merger (Saio \& Jeffery 2002), a low-mass star leaving the AGB shortly after the first thermal pulse (Miller Bertolami \& Althaus 2006) or a diffusion induced nova (Miller Bertolami et al. 2011). 

  If we speculate that IC~4663 has a companion, then perhaps a close binary evolution channel may be responsible for the helium-rich composition. Yungelson et al. (2008) proposed such a scheme to explain helium-rich AM Canum Venaticorum (AM CVn) stars (Warner 1995; Solheim  2010). IC~4663 may be the ejected common-envelope at one of these stages making it a possible AM CVn progenitor. If IC~4663 were a pre-CV, then it could potentially evolve into a helium-rich nova after additional angular momentum loss (e.g. V445 Puppis, Woudt et al. 2009). Another possibility is that a common-envelope merger already took place and this would be consistent with the apparent lack of evidence for a close binary (Sect. \ref{sec:var}). More generally this would help explain the apparent lack of [WR] central stars in close binaries, despite many nebulae around [WR] central stars showing post-CE morphologies (Miszalski et al. 2009b).
  
   The discovery of the [WN3] central star of IC~4663 has several implications. 
   Joining the O(He) stars (M\'endez 1991; Rauch et al. 1994, 1996, 1998, 2008) it is now the fifth hot ($T_\mathrm{eff}\ga100$ kK), He-rich and H-deficient post-AGB star known. A second post-AGB H-deficient sequence [WN]$\to$O(He) now seems to be a logical conclusion based on the strong match between their compositions. This has been hypothesised before (e.g. Werner 2012), but only now does it seem to be real with the [WN3] discovery in IC~4663. The RCB stars may also play an uncertain role in this sequence as RCB$\to$[WN]$\to$O(He). The nebula evolutionary status of IC~4663 and the two PNe with O(He) central stars (Sect. \ref{sec:nebevol}), being coincident with [WO] and PG1159 stars, respectively, supports the existence of a second sequence as their evolutionary status mirrors the [WC]$\to$PG1159 sequence. Once the stellar wind of the [WN3] central star of IC~4663 disappears it will likely become an O(He) central star and later a DO WD.

   In summary, we propose the [WN]$\to$O(He) sequence is a logical conclusion from the strong match between the surface compositions of the [WN3] central star of IC~4663 and the O(He) stars (Tab. \ref{tab:comp}). The evolutionary status of their nebulae (Sect. \ref{sec:nebevol}) suggests it forms a second parallel H-deficient evolutionary sequence to the existing carbon-rich [WC]$\to$PG1159 sequence. The relatively large amount of residual hydrogen in PB~8 suggests it may belong to a separate [WN] sequence to IC~4663, perhaps reflecting the dichotomy between H-deficient early-type WN stars and H-rich late-type WN stars (e.g. Hamann et al. 2006).

   \begin{table*}
      \centering
      \caption{Stellar compositions expressed as mass fractions (\%) of IC~4663 and other objects with similar compositions.}
      \label{tab:comp}
      \begin{tabular}{lllllllllllll}
         \hline
         Object & Type & $T_\mathrm{eff}$ & log $g$ & $\dot{M}$ & H & He & C & N & O & Ne & Reference\\
         &      &  (kK)            & (cgs)   & ($M_\odot$ yr$^{-1}$) &  &    &   &   &   &   & \\  
         \hline
         IC~4663 & [WN3] & 140            & 6.1$\dag$   &  $1.8\times10^{-8}$  & $<$2 & 95 & $<$0.1: & 0.8 & 0.05 & 0.2 & this work\\
         K~1-27 & O(He) & 105             &  6.5 &  $7.9\times10^{-10}$  &   $<$5 & 98 & $<$1.5 & 1.7 & - & - & Rauch et al. (1998)\\  
         LoTr~4 & O(He) & 120             & 5.5  & $2.0\times10^{-8}$ & 11 & 89 & $<$1 & 0.3 & $<$3 & - &   Rauch et al. (1998)\\
         PB~8  & [WN6/WC7] & 52           & - & $8.5\times10^{-8}$ & 40 & 55 & 1.3 & 2 & 1.3 & - & Todt et al. (2010a)\\ 
         RY Sgr & RCB &    7.25 & 0.7 & -& 6$\times$10$^{-4}$ & 98 & 0.7 & 0.3 & 0.09 & - & Asplund et al. (2000)\\
         \hline
      \end{tabular}
      \begin{flushleft}
         $\dag$ Adopted value assuming $M=0.6$ $M_\odot$.\\
         : Model constrained (no carbon lines observed).
      \end{flushleft}
   \end{table*}
  
\section{Conclusions}
\label{sec:conclusion}
We have discovered the first bona-fide [WN] central star of a PN in IC~4663. Our main conclusions are as follows:
\begin{itemize}
   \item The Gemini South GMOS spectrum of the central star is dominated by broad He~II and N~V emission lines, that together with the notable absence of He~I and C~IV, corresponds to a [WN3] subtype. 
   \item A \textsc{cmfgen} NLTE model atmosphere of the central star was built and reasonably reproduces the GMOS spectrum. The central star has a very hot ($T_\ast=140$ kK) and fast ($v_\infty=1900$ km s$^{-1}$) stellar wind. The atmosphere consists of helium (95\%), hydrogen ($<2$\%), nitrogen (0.8\%), neon (0.2\%) and oxygen (0.05\%) by mass. Such a composition cannot be explained by extant scenarios that seek to explain the H-deficiency of post-AGB stars.
   \item The helium-rich atmospheric composition strongly links [WN] central stars to the O(He) central stars (Rauch et al. 1998). The data support a second post-AGB H-deficient evolutionary sequence [WN]$\to$O(He) parallel to the established carbon-rich [WCL]$\to$[WCE]$\to$PG1159 sequence. Connections to other He-rich/H-deficient stars may include RCB stars, and if there were a binary companion to the [WN3] star, AM CVn stars and He-rich novae after further angular momentum loss. The existence of this second sequence strongly suggests another mechanism exists for producing H-deficient post-AGB stars besides the various born-again scenarios.
   \item For all reasonable distances the luminosity of the central star is 4--6 mag fainter than massive WN3 stars (Crowther et al. 1995b). At our assumed distance of 3.5 kpc (SSV08), the derived luminosity is log $L/L_\odot=3.6$, the stellar radius $R_\ast=0.11$ $R_\odot$ and the mass loss rate $\dot{M}=1.8\times10^{-8}$ $M_\odot$ yr$^{-1}$. The transformed radius, $R_t$, of 13.4 $R_\odot$ for IC~4663 is comparable to massive WN3 stars, indicating a similar stellar wind density to its massive counterparts. This favours our morphological [WN3] classification of the central star of IC~4663, rather than alternative weak wind (Of or Of-WR) classifications (M\'endez et al. 1990; M\'endez 1991).
   \item The surrounding nebula is a bona-fide PN. The elliptical main nebula has a relatively low expansion velocity of 30 km s$^{-1}$ and is surrounded by a newly discovered AGB halo. The evolutionary status of the highly ionised nebula corresponds to that of similar nebulae around [WO] type central stars and supports the [WN]$\to$O(He) evolutionary sequence. 
   \item Chemical abundances of the nebula were derived from a combination of \emph{IUE}, GMOS and CCM10 emission line intensities. The abundance pattern is approximately Solar with slight enhancements for helium, nitrogen and neon. The carbon to oxygen ratio is lower than in the Sun, however the data are insufficient to prove a carbon underabundance.
   \item We found no evidence for RV variability from the central star $\ga$5--10 km s$^{-1}$ in three epochs. Further monitoring of the central star should be performed to conclusively rule out the presence of a companion to the [WN3] central star.
\end{itemize}

\section*{Acknowledgments}
We acknowledge the conscientious and helpful support of Michelle Edwards, Henry Lee and Andrew Gosling during the course of our Gemini programme. 
   We thank S{\l}awomir G\'orny for valuable help in calculating $S_V$ and $S_{\mathrm{H}\beta}$ and Joanna Miko{\l}ajewska for discussions on symbiotic novae.
   AFJM is grateful to NSERC (Canada) and FQRNT (Quebec) for financial assistance.
    This work was primarily based on observations obtained at the Gemini Observatory, which is operated by the Association of Universities for Research in Astronomy, Inc., under a cooperative agreement with the NSF on behalf of the Gemini partnership: the National Science Foundation (United States), the Science and Technology Facilities Council (United Kingdom), the National Research Council (Canada), CONICYT (Chile), the Australian Research Council (Australia), Minist\'erio da Ci\^encia e Tecnologia (Brazil) and Ministerio de Ciencia, Tecnolog\'ia e Innovaci\'on Productiva (Argentina). It also used observations made with the NASA/ESA Hubble Space Telescope, and obtained from the Hubble Legacy Archive, which is a collaboration between the Space Telescope Science Institute (STScI/NASA), the Space Telescope European Coordinating Facility (ST-ECF/ESA) and the Canadian Astronomy Data Centre (CADC/NRC/CSA).

\label{lastpage}

\end{document}